\documentclass{ifacconf}
\usepackage{graphicx}
\usepackage{natbib}   
\usepackage{url}     
\usepackage{bm}
\usepackage{booktabs}       
\usepackage{amsmath}
\usepackage{amssymb} 

\usepackage[normalem]{ulem}

\usepackage{multicol}
\usepackage{multirow}

\usepackage{nicefrac}       
\usepackage{microtype}      
\usepackage{color}
\usepackage{xcolor}

\begin{document}
\begin{frontmatter}
\title{Parameter estimation in ODE models with certified polynomial system solving \thanksref{footnoteinfo}}

\thanks[footnoteinfo]{This work was partially supported by the NSF under grants CCF-2212460 and DMS-1853650.}

\author[AD]{Alexander Demin}
\author[AO]{Alexey Ovchinnikov}
\author[FR]{Fabrice Rouillier}

\address[AD]{HSE University, Moscow, Russia (e-mail: asdemin\_2@edu.hse.ru).}
\address[AO]{Department of Mathematics, CUNY Queens College, and Ph.D. Programs is Mathematics and Computer Science, CUNY Graduate Center, New York, NY, USA (e-mail: aovchinnikov@qc.cuny.edu)}
\address[FR]{Sorbonne Universit\'e, Paris Universit\'e,  CNRS (IMJ-PRG), Inria Paris, Paris, France (e-mail: Fabrice.Rouillier@inria.fr)}

\begin{abstract}
We consider dynamical models given by rational ODE systems. Parameter estimation is an important and challenging task of recovering parameter values from observed data. Recently, a method based on differential algebra and rational interpolation was proposed to express parameter estimation in terms of polynomial system solving. Typically, polynomial system solving is a bottleneck, hence the choice of the polynomial solver is crucial.
In this contribution, we compare two polynomial system solvers applied to parameter estimation: homotopy continuation solver from {\tt HomotopyContinuation.jl} and our new implementation of a certified solver based on rational univariate representation (RUR) and real root isolation. We show how the new RUR solver can tackle examples that are out of reach for the homotopy methods and vice versa.
\end{abstract}


\end{frontmatter}

\section{Introduction}
ODE models  are integral to scientific processes across many disciplines.  Model parameter values are required for analyzing the behavior of solutions. Computing these values from observed data is  a {\em parameter estimation} problem and has applications in areas ranging from epidemiology to chemical reaction networks and pharmacokinetics.

The state of the art for parameter estimation in ODE models is mainly composed of optimization-based and algebra-based approaches. For the former, even if the convergence can be proven, it is not known yet how to develop stopping criteria that would find the parameter values within the user-specified local error (see among many others e.g. \citep{AMIGO2} and the references given there).~
Potentially more robust, algebra-based approaches, tackle ODE models by exploiting theoretical results from differential  algebra, see \citep{PE} and the references given there {\color{black}for a comparison with optimization-based methods}. 

In this paper, we propose to tackle the efficiency bottleneck of a differential-algebra based approach~\citep{PE}. A key step in this algorithm is finding all solutions of a polynomial system constructed from the ODE model and data. In the current implementation, polynomial solving is done via the technique of homotopy continuation. Being useful in many cases, it can be not as efficient and accurate as needed. We have discovered that such polynomial systems typically have very few solutions and that certified polynomial system solver that use Gr\"obner basis and rational univariate representation (RUR) can be not only more reliable but more efficient too.

\section{Main result}

We begin with the problem statement \citep{PE}.

\noindent\textbf{Input:}
 An ODE model $\Sigma$
        \begin{equation}\label{eq:main}
          \begin{cases}
            \bm{x}'(t) = \bm{f}(\bm{x}(t),  \bm{u}(t),\bm{\mu}), \\
            \bm{y}(t)\ = \bm{g}(\bm{x}(t),  \bm{u}(t),\bm{\mu}),  \\
            \bm{x}(0)\, = \bm{x_0},
          \end{cases}
        \end{equation}
        where we use bold fonts for vectors $\bm{f}$ and $\bm{g}$ of
                rational functions describing the model,
        $\bm{x}$ vector of state variables,
         $\bm{u}$ vector of input (control) variables, which are known,
          vector $\bm{y}$ of output variables, and
          vectors $\bm{\mu}$ and $\bm{x_0}$ of unknown  parameters; and
        
Data $D=((t_1,\bm{y}_1),\ldots,(t_n,\bm{y}_n))$, where $\bm{y}_i$ is the measured value of~$\bm{y}$ at time~$t_i$.

\noindent\textbf{Output:} Estimated values for the parameters $\bm{\mu}$ and $\bm{x_0}$.

Consider the toy example from~\citep{PE}
with
\[
\begin{gathered}
\Sigma = 
\begin{cases}

  \quad\,    x'        =  -\mu x \\
  \quad\ \,    y                  =  x^2+x \\
      x\left(  0\right) =  x_{0} 
    \end{cases}\\
D =
  \{(0.00, 2.00),(0.33,1.56), (0.66,1.23),
   (1.00, 0.97) \}.
   \end{gathered}
   \]
{\color{black} The approach from~\citep{PE} produces the following (polynomial) system}
{\color{black}($x_0,x_1,x_2$ represent $x,x', x''$)}:
   \[\begin{cases}
                    \ \  2.00  {\color{black}=}  x_0^2+x_0,                        \\
                    -1.50  {\color{black}=}  2x_1x_0 + x_1,                    \\
                   \ \ \, 1.22   {\color{black}=}  2(x_1{\color{black}^2} + x_0x_2)+x_2,\\
                 \quad\ \    x_1    =       -\mu x_0.     \end{cases}
                \]
This is a system of four degree  $2$ equations in $4$ variables. Such systems typically have $2^4 = 16$ solutions. However, this system only has $2$ solutions, with the values of $(\mu,x_0)$ being $(0.49,1.00)$ and $(0.25,-2.00)$. This example is a heuristic illustration of a much lower than expected number of solutions of the polynomial systems we are working with. {\color{black} A rigorous analysis is left for future research.}

\begin{table}[h]
\centering
\caption{\small The running time (in seconds) and maximal relative errors (in percentage) of estimated {\color{black}parameters} using backends RUR (New) and HC Julia, n/a means no result, {\color{black}OOM} means out of memory ($>100$ GB), timeout means estimation took more than 1 day.}
\small
\begin{tabular}{lrrrrrr}
\multirow{2}{*}{Model} & \multicolumn{2}{c}{Model data} & \multicolumn{2}{c}{RUR (New)} & \multicolumn{2}{c}{HC Julia} \\ 
& States & Params. & Time  & Error & Time & Error\\
\hline
Akt-1 & 9 & 9 & $900$ & 0.0  & $1800$ & n/a \\
Akt-2 & 9 & 17 & $3600$ & 10.0  & timeout  & n/a \\
\hline
Crauste-1 & 5 & 13 & $1$  & 0.2 & $4$  & $0.0$  \\
Crauste-2 & 5 & 13 & $24$ & 0.2  & $17$  & $0.0$  \\
Crauste-3 & 5 & 13 & $768$  & 7.0 & $57$  & $0.0$ \\
\hline
NFkB-1 & 16 & 5 & $10$  & $0.0$  & $80$  & $0.0$  \\
NFkB-2 & 16 & 15 & {\color{black}OOM} & n/a & timeout & n/a \\
\hline
Goodwin & 3 & 7 & $1$  & $0.0$  & $1$  & n/a \\
Treatment & 4 & 5 & $1$ & $0.0$ & $15$ & n/a\\
PK1 & 4 & 10 & $1$ & $0.0$ & $7$ & n/a \\
CRN & 6 & 6 & $1$  & $5.0$  & $240$  & $5.0$  \\
SEIR 36 & 10 & 11 & $10$ & $5.0$ & $300$ & $5.0$ \\
\end{tabular}
\label{table:data}
\end{table}

Larger ODE models lead to larger polynomial systems, where finding solutions becomes a challenge on its own. 
However, all models we tried shared the same property: relatively small number of solutions of the polynomial systems we construct from Lie derivatives.
For solving such polynomial systems, the original implementation from~\citep{PE} used the\linebreak{\tt HomotopyContinuation.jl} package in Julia as the default choice~\citep{HomotopyContinuation.jl}.

We report on our experience using the new solver based on rational univariate representation and root isolation implemented in {\tt RationalUnivariateRepresentation.jl} and {\tt RS.jl}~\citep{demin:2024}. We compare the new RUR solver with the current {\tt HomotopyContinuation.jl} solver (HC Julia).
Our benchmark includes dynamical models of varying sizes from~\citep{Villaverde2023}. 
We use the Julia language running on i9-13900 CPU.

Table \ref{table:data} summarizes our findings\footnote{For each benchmark model, we provide the code to reproduce our findings, available at: \url{https://github.com/iliailmer/ParameterEstimation.jl/tree/main/rur-and-hc}.}. For each model in the table, we report the number of states and parameters in the model, the running time, and the relative error of estimation obtained using the new RUR solver and the HC Julia solver. {\color{black} We tested one set of parameters per model, because we expect the solvers to behave similarly for different numerical values.}
Examples roughly fall into different groups:

\begin{itemize}
    \item Out of reach for HC Julia but solvable with RUR: Akt-2 model. The associated polynomial system was solved in about 1 hour with RUR, but did not finish in a day with HC Julia. The system has 69 unknowns; the B\'ezout bound for the number of solutions is $2^{26} \cdot 3^{37}$, which would have been hopeless for any algebraic solver; luckily, the actual number of solutions is $80$. 
    
    \item No solutions returned by HC Julia but RUR found solutions: Akt-1, Goodwin, PK1, and Treatment models. A common feature of these models is the presence of structurally non-identifiable states or parameters{\color{black}~\citep{Villaverde2023}}. Non-identifiability may cause certain solution coordinates to blow up and make numerical solving unstable.

    \item More challenging for RUR: Crauste-2, Crauste-3 models. Crauste-1 is a model of the behavior of CD8 T-cells introduced in \citep{crauste}; both RUR and HC Julia readily solve it. We would expect homothopy continuation methods to be efficient on a large class of examples but it turned out that we had difficulties to illustrate that with our small set of examples. Thus, we construct Crauste-2 and Crauste-3 artificially from Crauste-1 by introducing symmetries in the model by squaring some parameters in the equations.
\begin{table}[h]
\centering
\caption{\small The data on the polynomial systems produced in parameter estimation task in the Crauste series.}
\small
\begin{tabular}{lrrr}
Model & Unknowns & Solutions & B{\'e}zout bound\\
\hline
Crauste-1 & 43 & 32 & $2^{5} \cdot 3^{20}$  \\
Crauste-2 & 43 & 128 & $2^{13} \cdot 3^{16}$ \\
Crauste-3 & 43 & 512 & $2^{23} \cdot 3^{11}$ \\
\end{tabular}
\label{table:crauste}
\end{table}
Although the number of unknowns remains unchanged, from Table \ref{table:crauste} we see that the structure of polynomial system changes and the number of solutions increases (both actual and in the theoretical upper bound, called B\'ezout bound). Typically, this makes solution via RUR harder.

\item Solved by both solvers: Crauste-1, NFkB-1, CRN, and SEIR 36 model{\color{black}s}.

\item Out of reach for both solvers: NFkB-2 model. The associated polynomial system has 122 indeterminates.
\end{itemize}

In conclusion, we see that our new algebraic solver can be competitive in performance and more robust on a some not cherry-picked parameter estimation problems with models roughly up to 10 states \& 10 parameters.

\bibliography{main}

\end{document}